**Machine learning-assisted cross-domain prediction of ionic conductivity in sodium and lithium-based superionic conductors using facile descriptors.**

## Abstract

Solid state lithium- and sodium-ion batteries utilize solid ionicly conducting compounds as electrolytes. However, the ionic conductivity of such materials tends to be lower than their liquid counterparts, necessitating research efforts into finding suitable alternatives. The process of electrolyte screening is often based on a mixture of domain expertise and trial-and-error, both of which are time and resource-intensive.

Data-driven and machine learning approaches have recently come to the fore to accelerate learnings towards discovery. In this work, we present a simple machine-learning based approach to predict the ionic conductivity of sodium and lithium-based SICON compounds. Using primarily theoretical elemental feature descriptors derivable from tabulated information on the unit cell and the atomic properties of the components of a target compound on a limited dataset of 70 NASICON-examples, we have designed a logistic regression-based model capable of distinguishing between poor and good superionic conductors with a cross-validation accuracy of over 82%. Moreover, we demonstrate how such a system is capable of cross-domain classification on lithium-based examples at the same accuracy, despite being introduced to zero lithium-based compounds during training. Through a systematic permutation-based evaluation process, we reduced the number of considered features from 47 to 7, reduction of over 83%, while simultaneously improving model performance. The contributions of different electronic and structural features to overall ionic conductivity is also discussed, and contrasted with accepted theories in literature. Our results demonstrate the utility of such a simple, yet interpretable tool provides opportunities for initial screening of potential candidates as solid-state electrolytes through the use of existing data. Given enough data utilizing suitable descriptors, high accurate cross-domain classifiers could be created for experimentalists, improving laboratory and computational efficiency.

Yijie Xu[a,b]

Yun Zong[b*]

Kedar Hippalgaonkar[b,c*]

[a] *Department of Chemistry, University College London, 20 Gordon Street, London WC1H 0AJ, United Kingdom*

[b] *Institute of Materials Research and Engineering (IMRE), A*STAR Agency for Science, Technology and Research, 2 Fusionopolis Way, Innovis #08-03, 138634, Singapore*

[c] *School of Material Science and Engineering, Nanyang Technological University, 639798, Singapore*

*Email correspondence to: y-zong@imre.a-star.edu.sg, kedar@ntu.edu.sg

## Introduction

Lithium ion batteries (LIBs) have become the forefront of energy storage technology for a wide range of applications, including mobile devices and electric vehicles. Similarly, Na-ion batteries have shown great promise as being a safer, more cost-effective alternative for larger grid-storage applications. However, both systems suffer from well publicized stability and safety issues.[1, 2] One of the these originates in their use of liquid electrolytes consisting of an ionic salt dissolved in an organic solvent, such as $LiPF_6$ or $NaPF_6$ in ethylene and dimethyl carbonate. While such electrolytes possess high ionic conductivity while remaining affordable, their inherent flammability increase the risk of catastrophic failure scenarios through mechanical damage, side reactions, and pressure build-up originating from detrimental thermal runaway scenarios.[3]

In contrast, solid ionic electrolytes offer higher energy densities while improving operational safety by reducing the overall flammability and limiting the prevalence of side reactions and consequently allowing for the use of higher voltage cathodes. One of the primary issues in their development has been their lower ionic conductivities when compared to their liquid counterparts, often in the scale of several orders of magnitudes.[4]

NASICON compounds refer to materials of the general simplified formula $NaMP_3O_{12}$, consisting of a skeleton made up of corner-sharing $MO_6$ octahedra and $PO_4$ tetrahedra, with channels accommodating the mobile Na cations.[5] First demonstrated as being suitable sodium-based solid electrolyte materials by Lalere et al in 2014, they have been well characterized and studied since the 1970s, resulting in a wealth of ionic conductivity data being available for analysis.[6]

However, the development of suitable candidates has traditionally required either trial-and-error or systematic approaches. While the former approach is possible,[7] both approaches would be prohibitively resource and labour-intensive.[8] To accelerate this search process, data-driven approaches utilizing machine learning have rapidly become more popular in modelling desirable properties in chemistry, covering a variety of applications including the prediction of ternary oxide structures,[9] identification of new high temperature piezoelectrics,[10] prediction of compound thermodynamic stability,[11] screening of new catalyst materials,[12] optimization of hydrothermal synthesis methods,[13] the design of new electrolyte materials,[14] and the prediction of reaction productions in organic synthesis.[15]

These machine learning approaches share the commonality in their attempt to capture some universal and possibly non-obvious information using certain feature-based representation that can be used to correlate to a particular desired material property for a targeted application. In their seminal work, Raccuglia *et al.* converted existing knowledge about failed and successful hydrothermal experiments into calculated cheminformatical features together and tabulated atomic properties in order to predict successful conditions for vanadium selenite crystallization.[13] Most importantly, they possess a capability to extrapolate beyond their training domains in order to predict properties of foreign examples not found within their training dataset, which is achieved by feeding the model a diverse set of training examples.[15] In the case of ionic conductivity of solid-state electrolytes, a suitable set of easily computed or measured material descriptors or features may be obtained that allow for the generation of a predictive model capable of guiding future research direction with only limited available data at a rate consistently shown to be better than random trial-and-error.

In addition to allowing for data-based evaluation of literature hypotheses on the strongest factors related to ionic conduction, a machine-learning approach is significantly more robust against human bias when evaluating the potential of different compounds. In this study, we've chosen to utilize a relatively simple data-driven approach relying on a predictor model trained on primarily theoretically-derived elemental and structural features, as the electronic structure-based calculations of ionic conductivity often require computationally expensive and complex calculations, which are undesirable for simple screening initiatives.

**Training Set**

A training dataset consisting of 70 experimentally synthesized NASICON materials of R3c symmetry was collected from literature to generate a feature set containing experimentally measured ionic conductivity values, along with simple molecular descriptors, structural descriptors, and electronic descriptors.[5, 16-34] The separation of these descriptors was done by category, and is shown below in Table 1.

Table 1. Categorization of descriptors used in study.

| Category | Examples | Count |
|---|---|---|
| Simple molecular descriptors | Formulaic species count, numeric species occupancy | 23 |
| Structural descriptors | Unit cell parameters, unit cell volume, ionic radii | 18 |
| Electronic descriptors | Electronegativity, ionic conductivity | 8 |

The formulas of our NASICON materials and accompanying ionic conductivities are listed in Table S1a and S1b, while the methods of determination for each descriptor (henceforth feature) are listed in Table S1c. A distribution of ionic conductivities along with the logarithmically transformed values is shown in Figure 1. The measured ionic conductivity values of our dataset ranged from $10^{-3}$ to $10^{-13}$ S $cm^{-1}$ in magnitude, although a strong over-representation is observed for lower ionicly conducting examples, particularly of those conducting less than $10^{-8}$ S $cm^{-1}$.

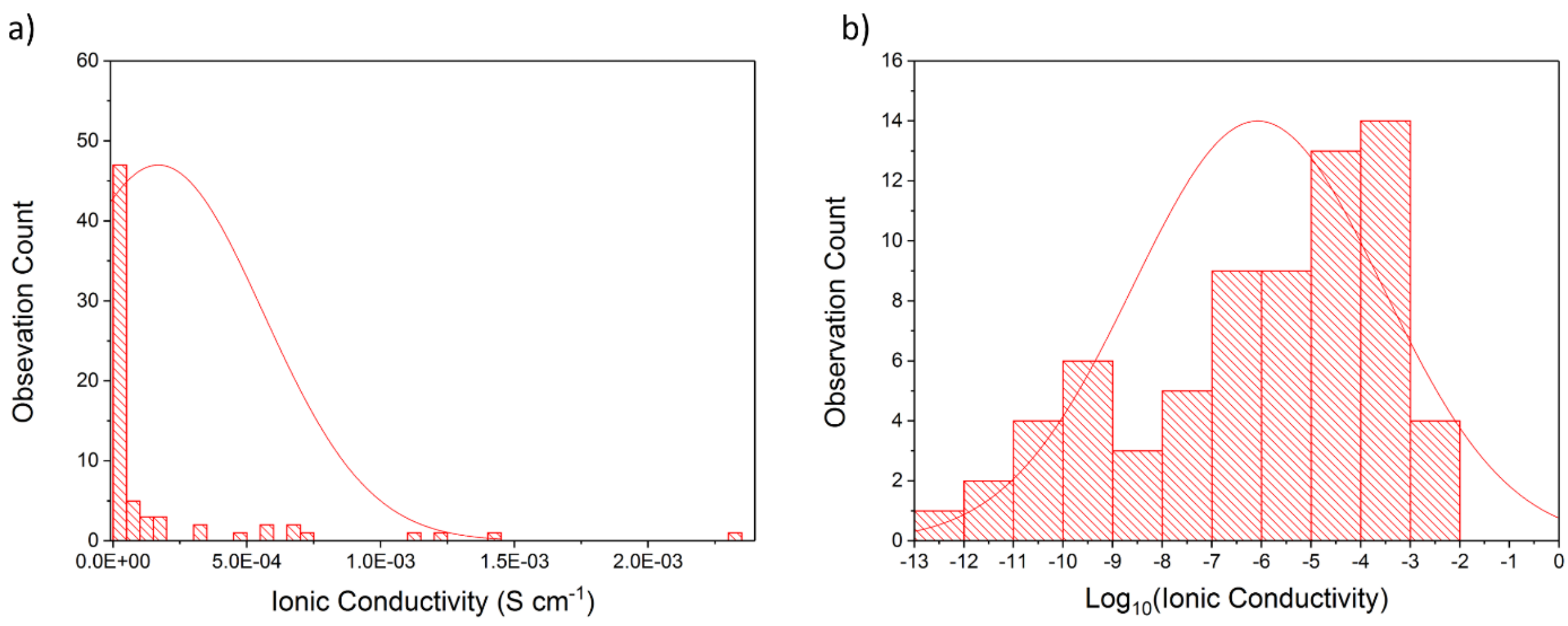

**Figure 1.** Distribution of the ionic conductivity of our training data (a) used in our model, and logarithmic plot of the former (b). Bin size was kept at $5 \times 10^{-5}$ S cm$^{-1}$ and 1, respectively.

The decision to include poorly conductive labelled examples was primarily due to both necessity to obtain data, and supported by the work by Raccuglia *et al.* , who demonstrated the importance of exposing any machine learning model to negative examples in order to contribute to more accurate predictions.[13] However, the poor representation of highly conducting examples in the dataset may influence the performance of our model when classifying such examples.

Besides the experimentally reported ionic conductivities,  the dataset was augmented with a subset of features that may or may not be correlated to the ionic conductivity. A full list of these features together with their descriptions is exhibited in Table S1b, along with their respective Pearson correlations with experimentally measured ionic conductivity. As Pearson correlation coefficients measure the degree of linear correlation between two values, a high correlation with our desired target feature of ionic conductivity could artificially increase our classification accuracy. These features were derived from a combination of unit cell parameters and elemental descriptors derived from tabulated information, chosen based on domain expertise as well as literature suggestions, and targeted compositional as well as physical and electronic information of our compounds.[5] While the Inorganic Crystal Structure Database was consulted for detailed crystal structure information on the compounds as a possible model feature, this was both sparse and incomplete for our dataset, and were hence not considered in our study.[35]

**Model selection**

To build an ionic conductivity predictor using our dataset, a Logistic Regression classification model was implemented. The motivation for the use of logistic regression is given in Supplementary Information S3**.** Briefly, logistic regression relies upon the use of a sigmoid function of the form shown below in Equation 1:

(1) $$P_{LogReg}(\tilde{\sigma}^{i} = 1) = \left[1 + \exp\left(-\sum_{j=0}^{n} \theta_j x_j^i\right)\right]^{-1}$$

Where $\theta_j$ refers to the regression coefficients of each feature $x_j$, and $n$ is the total number of features. The sigmoid function component of the logistic regression equation converts the raw scalar value into one between 0 and 1, representing that probability that the particular material belonging to class 1, or excellent ionic conductor. The final predicted output of the logistic regression model is dependent on the class with greater likelihood, i.e. $\tilde{\sigma}^{(i)} = 1$ if the material *i* is predicted to be an excellent ionic conductor (P>= 50%), and $\tilde{\sigma}^{(i)} = 0$ if the material *i* if the material is predicted to be a poor ionic conductor (P< 50%).

We established this decision boundary of $1 \times 10^{-6}$ S $cm^{-1}$ by examining our dataset, to ensure an approximate equal numerical representation across the two classes. In support of this, ionic conductivity values above the boundary are classified as good ionic conductors (with a label of 1), and those below classified as poor ionic conductors (with a label of 0). Using our criteria, approximately 42.9% of our dataset were considered class 0 and 57.1% considered class 1. Alternative decision boundaries were explored, but found to yield significantly more unbalanced output classes. Imbalanced class distributions in datasets have been well studied for classification applications and have been shown to be linked to poorer classification performance.[36]

**Feature Selection**

47 different computed or experimentally measured features were collected for both the training and test datasets of NASICON structures. A complete list of considered features can be found in Table S1b. These carefully selected features encompass both structural and electronic attributes of their compounds, and were chosen for their computational simplicity, wherein every attribute could be derived from tabulated information associated with its unit formula.

To build a primarily theoretically-based predictive model free from the need of experimental inputs, experimentally-measured features such as activation energies were excluded from the feature list of the final model. Features related to specific elements in the test set, such as the number of a constituent atom not found within the training dataset, were excluded in order to build predictive robustness, prevent overfitting, and to evaluate classification accuracy on test examples featuring foreign elements.

The final model features used in the modelling process, together with their individual Pearson correlations with ionic conductivity for the training dataset, are given in Table S1.

**Evaluation**

The pipeline for our evaluation process is shown below in Figure 2.

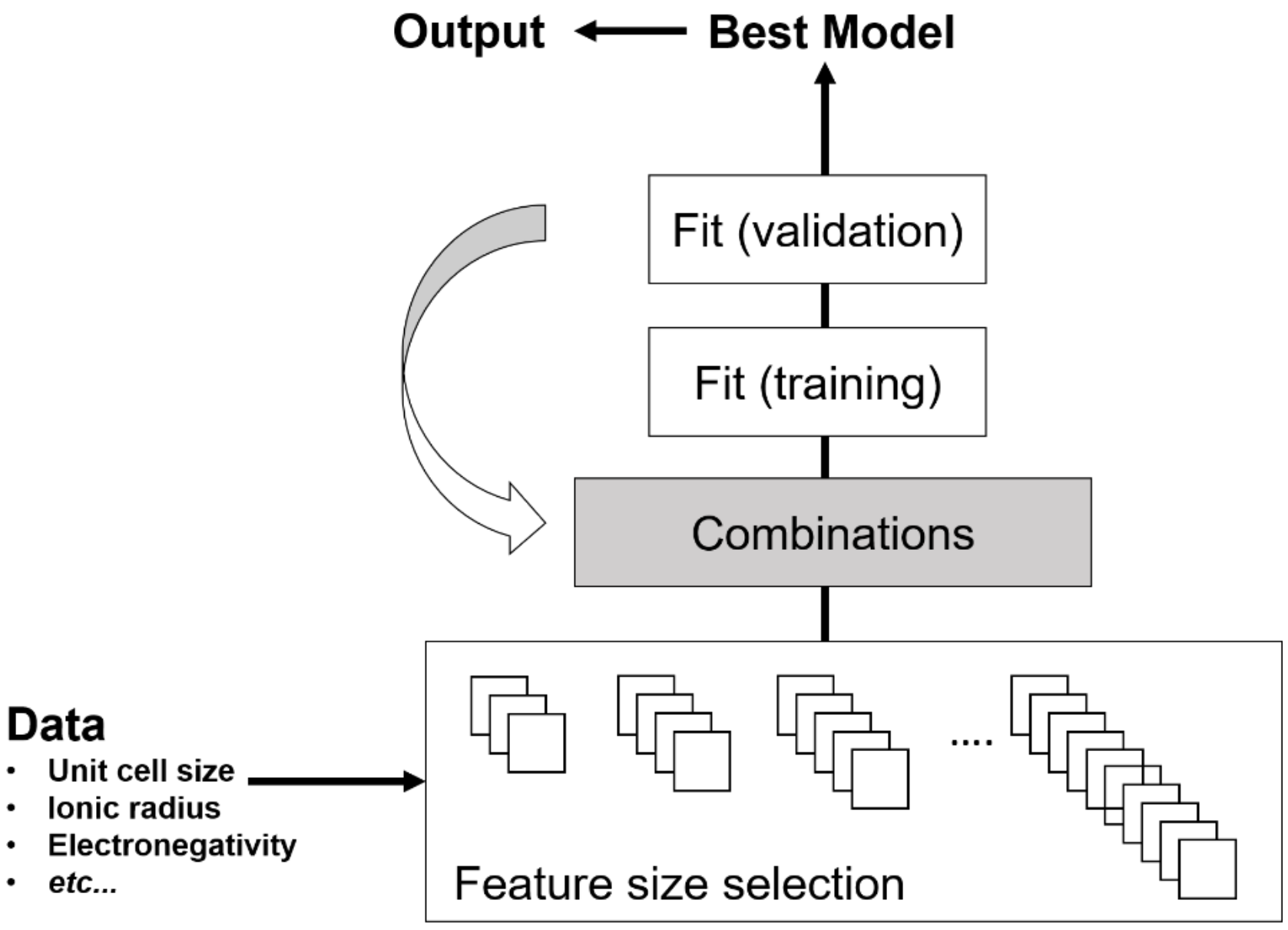

**Figure 2.** Model evaluation process used within our study. Combinations of different-sized feature sets are systematically evaluated for training and validation accuracies, with the best performing combination at each feature size recorded.

To identify a model with the best-performing feature combination a systematic evaluation of the features was performed as for each *n* -sized feature set (where *n* is between 3 and 10, inclusive): a logistic regression model was constructed using all possible permutated combinations of features for that fixed feature set size. The number of features used to fit the data was chosen in accordance to literature evidence of the use of keeping the minimum 5:1 ratio between data points and the number of features to avoid overfitting and promote good generalization capability.[37] Although it is considered standard practice to split a dataset into training and validation subsets at a ratio of roughly 80:20, the small size of our dataset necessitated the use of fold-based cross-validation approaches.

For a specific feature set size, the training dataset was fitted to their respective ionic conductivity class labels utilizing the permutated subset of features and the resulting training accuracy (TAC), or the difference between the predicted and true outputs, was calculated . In the context of machine learning, a model possessing an accuracy of 0.7 would mean that given 100 examples, an average of 70 would be correctly classified. This is given by Equation 2:

(2)

$$TAC = \frac{1}{M}\sum_{i=1}^{M}\left(\tilde{\sigma}^i = \hat{\sigma}^i\right)$$

Where $\tilde{\sigma}^i$and $\hat{\sigma}^i$ refer to the predicted and true class of a training example from a M-sized dataset. Values of 1 and 0 were added into the sum for each correct and incorrect prediction, respectively, before the average was taken to obtain the accuracy value. From the TAC results of all models at the specified feature set size, models with the TAC of more than 0.8 were selected for cross-validation evaluation. Due to the small size of our dataset, the cross-validated accuracy (CVAC) was computed through K-fold stratified cross-validation (SCV), with a *K* value of 7. Briefly, 10 stratified datapoints were systematically removed as a validation set, with the model retrained on the remaining N-10 data points, following which the accuracy was measured on the validation subset. This was then repeated for the length of the entire dataset, with an overall mean accuracy and standard deviations recorded. This can be described by Equation 3 :

(3)

$$CVAC = \frac{1}{M}\sum_{i=1}^{M}\left(\tilde{\sigma}^i = \hat{\sigma}^i_{cv}\right)$$

Where $\tilde{\sigma}^i$and $\hat{\sigma}^i_{cv}$ refer to the predicted and true class of a cross-validation example from a M-sized dataset. Values of 1 and 0 were added into the sum for each correct and incorrect prediction, respectively, before the average was taken to obtain the accuracy value. The CVAC was used as a measure of predictive capacity of each model at the specified feature set size, giving a good summary of the level of under- or overfitting observed with each feature combination. Generally, models that exhibit high training accuracy together with lower cross-validation accuracy suggest that overfitting has taken place, as the model fails to generalize or extrapolate outside its training dataset. In contrast, models with low training accuracy together with low cross-validation accuracy suggest the presence of underfitting, or that the model lacks sufficient data or suitable features for prediction.

Following validation analysis, models of each feature set size with high CVAC scores and low standard deviations were evaluated using an unseen test set, consisting of 11 NASICON compounds from literature with documented ionic conductivities and 11 LISICON compounds added as an evaluation of the model's extrapolation capabilities.[38-50] These lithium compounds were  added due to the similar behavior of the two systems in electrochemical storage applications and as a test of the generalization capabilities of our model beyond its original training domain. The test accuracy or validation misclassification rate (VAC) was evaluated in the same manner as the TAC mentioned above. Since each set of cross-validation predictions was made on an experimental sample that was removed from the training set, the CVAC together with the VAC captures the potential predictive power of the model on unseen data and provides a way to evaluate its usefulness to researchers in the field.

## Results and Discussion

### Model Performance

The results of our evaluation process are displayed in Figure 3 and Table 2, respectively. Figure 3 displays the accuracies and standard deviation metrics of the best feature combination at a specified feature set size $n$, while Table 2 lists the exact feature combination that gives rise to the performance metrics specified in the Figure 2.

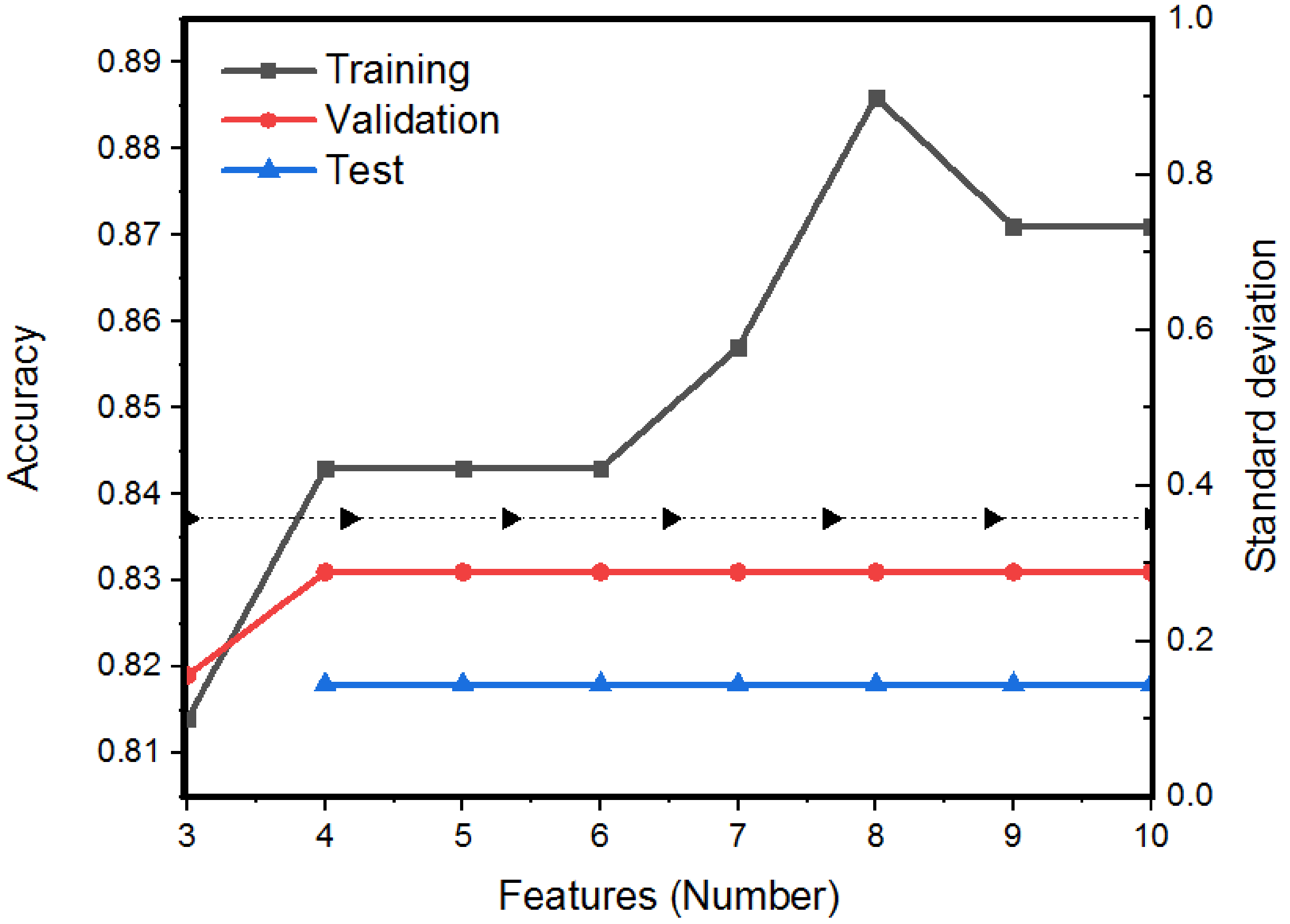

**Figure 3.** Results from systematic feature evaluation, displaying the best model performance across permutated feature sets of size *n*, where *n* is between 3 and 10, inclusive. Performance was evaluated by training accuracy (black), cross-validated accuracy (red), test accuracy (blue), and model standard deviation (black triangles with dotted line, right axis). Not pictured: a VAC of 32.8% was obtained for the best performing model using 3 features.

Table 2. Output accuracies of models trained with permutated feature combinations across different feature set sizes. Training and cross-validation accuracies are based on the training dataset, and validation accuracy is based on the test dataset. The bolded entry represents the final selected optimal model. Here, the symbols D1, D2, and D3 refer to the elements occupying the Na, M1, and M2 positions in $NaM1M2P_3O_{12}$. Detailed descriptions of the individual feature labels can be found in Table S1b.

| Feature set size | Feature combination | TAC | CVAC | Standard deviation | VAC |
|---|---|---|---|---|---|
| 10 | 'a', 'c', 'h', 'd', 'D1ionicr', 'D2ionicr', 'PO4', 'D1eneg', 'D2eneg', 'D3eneg' | 0.871 | 0.831 | 0.358 | 0.818 |
| 9 | 'a', 'c', 'h', 'd', 'D2ionicr', 'PO4', 'D1eneg', 'D2eneg', 'D3eneg' | 0.871 | 0.831 | 0.358 | 0.818 |
| 8 | 'a', 'c', 'h', 'd', 'D2ionicr', 'PO4', 'D2eneg', 'D3eneg'<br>**'c', 'h', 'd', 'D1ionicr', 'D2ionicr', 'PO4', 'D2eneg', 'D3eneg'** | 0.886<br>**0.871** | 0.831<br>**0.831** | 0.358<br>**0.358** | 0.818<br>**0.818** |
| 7 | 'c', 'h', 'd', 'D2ionicr', 'PO4', 'D2eneg', 'D3eneg' | 0.857 | 0.831 | 0.358 | 0.818 |
| 6 | 'a', 'a/c', 'D1ionicr', 'PO4', 'D2eneg', 'D3eneg' | 0.843 | 0.831 | 0.358 | 0.818 |
| 5 | 'a', 'D1ionicr', 'PO4', 'D2eneg', 'D3eneg' | 0.843 | 0.831 | 0.358 | 0.818 |
| 4 | 'a', 'PO4', 'D2eneg', 'D3eneg' | 0.843 | 0.831 | 0.358 | 0.818 |
| 3 | 'D3vol', 'Na', 'D3eneff' | 0.814 | 0.819 | 0.397 | 0.328 |

Overall, a maximum training accuracy of 0.886 was achieved with a feature size of 8. A high CVAC accuracy and VAC accuracies of 0.831 and 0.818 was achieved by all models featuring all feature sizes except those with feature size 3. This high similarity in accuracy levels between the majority of feature sizes can be explained by the small size of the validation and test sets, and it is expected that large sets would show greater discrepancies as a greater number of examples featuring a wider possible range of features would be evaluated.[51]

The positive correlation between training accuracy and feature size can be explained by the incorporation of more features such as species electronegativity, ionic radii and structural features such as lattice parameters, unit cell hypotenuses, which are able to capture information universally shared across all compounds. When the feature size was reduced to 3, the need to maximize training accuracy resulted in strong overfitting to the training dataset being observed using possibly underlying overrepresented features, as evidenced by the poor performance of the model on the test set. Interestingly, if the number of features used exceeded 8 ($n>8$), the training accuracy of the models decreased by 1.7%, which may be due to the incorporation of the sodium-specific features (namely the ionic radius and electronegativity) that are in conflict with existing feature sets, as the immutable nature of these features could cause them to behave as a constant bias value, shifting the existing fit in the y-axis.

Following the cross-validation process, the extrapolative performance of our models was evaluated using the test set mentioned in the previous section. From our results, we notice that the same validation accuracy is shared among most of our models, which can be explained by the relatively small test dataset. Interestingly, when considered separately, the test accuracy of our sodium and lithium test datasets remains the same, at roughly 83% each, demonstrating the extrapolation capabilities of our model when presented with previously unseen data of a similar domain. The similarity of lithium and sodium-ion conduction has been investigated extensively in literature for energy storage purposes, supporting the outputs of our model. [52] In literature, logistic regression models utilizing more complex computationally- derived features for lithium-ion based solid electrolyte screening have been investigated by Sendek *et al.* While the study achieved a higher cross-validation accuracy 90%, the feature complexity of the model demanded extensive computational effort, reducing its effectiveness in rapid screening. The capability of our model to capture much of the same performance using easily derivable elemental information with minimal data pre-processing validates its utility as a preliminary screening tool.

## Discussion

As all our models give the same levels of validation and test accuracies past $n=4$, the selection criteria for the final optimal model feature were high accuracy, minimal overfitting on existing training data, and maximal generalization capability. The relationship between the former two factors can be controlled by comparing CVAC with TAC, where a difference between training and validation accuracies would suggest a high level of overfitting. In our study, this difference was arbitrarily set at below 5%. Similarly, the generalization capability was judged by selecting models featuring features with a minimal amount of intra-category cross-correlation. Cross-correlation was measured through a cross-correlation matrix plotted on the features identified in Table S1 using the Python Seaborn statistical package.

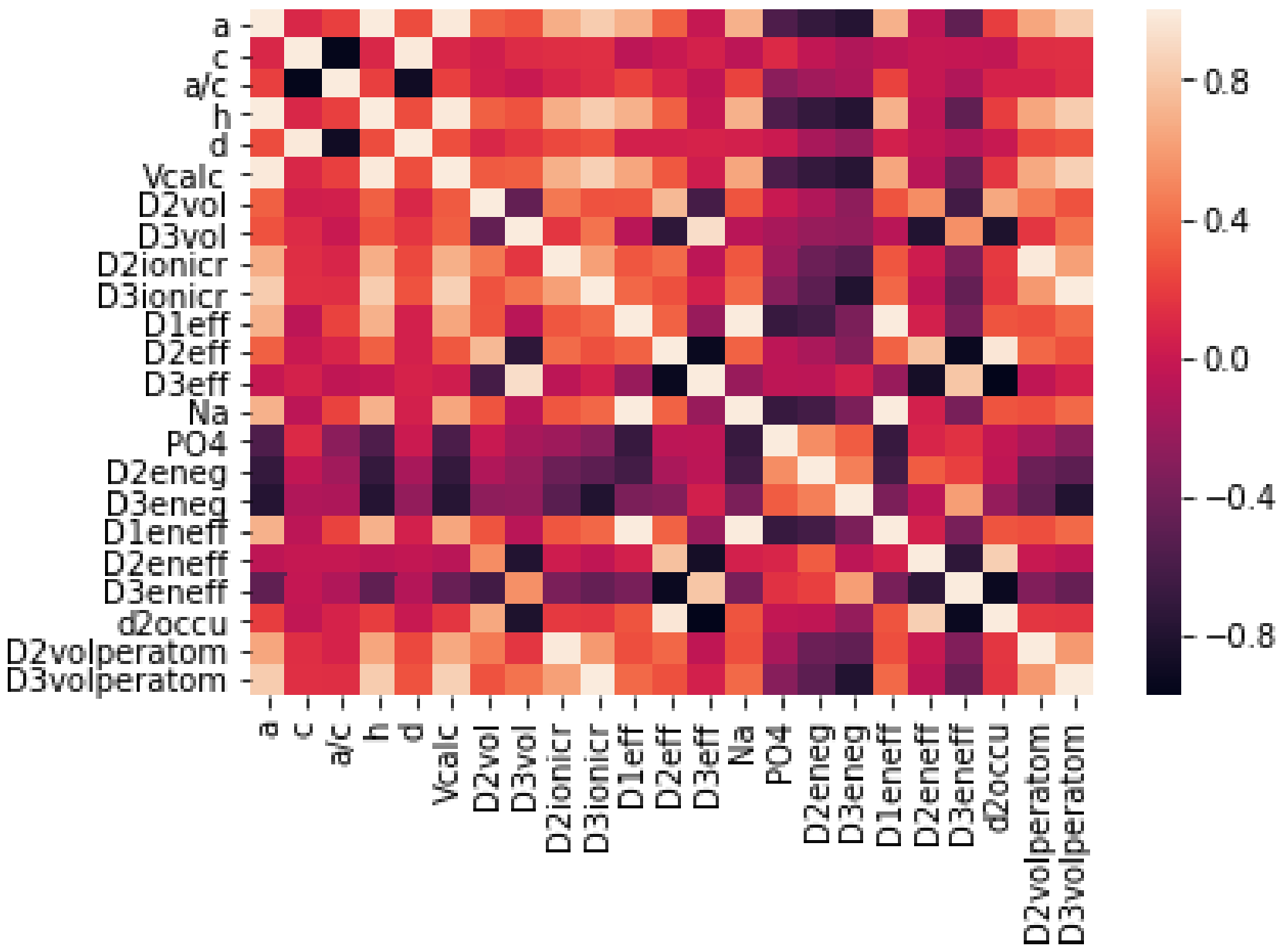

Figure 4. Correlation matrix of the features used within in this study (excluding features based on species element -specific counts). Degree of correlation is indicated by the hue of the image.

The results from Figure 4 demonstrate some explicit intra-category correlations between our features. For example, structural parameters tend to be positively correlated with each other and the overall volume of the unit cell. In particular, features *"a", "h",* and *"Vcalc"* were highly cross-correlated, which is understandable given their interdependence during calculations. Similarly, features specific to the cationic structural of the compound, such as the specific electronegative features *"D3eneff"* and *"D2eneff"*, also exhibit strong correlation due to a proportion of compounds exhibiting the same elements for M2 and M3 positions in $NaM1M2P_3O_{12}$. Furthermore, some of the observed correlations are significant in implication to warrant further study. For example, a high unit cell volume is shown to be correlated with a high number of sodium ions, lower count of phosphate groups and fewer electronegative structural cationic species, suggesting that bond ionicity is partially related to the volume of a unit cell.

As such, the selection of the final optimal model (bolded under $n=8$ in Table 1) was also based on a small VAC-TAC accuracy difference of 4.8%,a minimal the number of highly cross-correlated features, and a minimal number (1) of highly Pearson-correlated features. The coefficients for the selected features were then extracted for analysis.

Thus, the final Logistic Regression model can thus be described with Equation 4,

(4)

$$\sum_{i=0}^{n} \theta_i X_i = 0.017c + 0.519h + 0.195d + 0.076 Ionic\ radius_{D1} - 0.072 Ionic\ radius_{D2} - 0.928 N_{PO_4} - 0.726 Electronegativity_{D2} - 0.057 Electronegativity_{D3}$$

The nature of the coefficients in our equation gives some intuition behind the underlying elemental attributes that affect ionic conductivity and allows us to examine literature theories. Equation 4 suggests an increase in ionic conductivity with increasing number of structural parameters – in fact, an increase in the number of structural unit cell features was observed to increase the training accuracy of the model when $n>7$. As all our examples share the R3c rhombohedral spacegroup, it can be argued these lattice parameters act as a representation of the cell volume as suggested by our cross-correlation analysis. It must be mentioned that the lack of the unit cell volume feature itself in the final models suggests that structural parameters are capturing some information beyond the spacegroup's calculated volume.

In literature, the role of free unit cell volume in gauging ionic conductivity was first proposed by Sammels *et al.*, who theorized that a large difference between unit cell volume and the ionic volume of its constituents would lead to enhanced ionic conductivity, and concluded that a cubic or pseudo-cubic unit cell structure was preferable for such applications.[53] Experimentally, a positive correlation between unit cell volume and ionic conductivity has been observed for superionic selenide, sulfide, and oxide compounds by Ong *et al.* Interestingly, such a correlation was observed until an optimum size of an ionic channel is reached, after which further volume increases have negligible effect, although this is not seen in our analysis as feature information on the size of ionic channels was not incorporated.[54] These conclusions are supported by the positive coefficients of our structural parameters in Equation 4, which given the positive correlation of structural parameters with unit cell volume, support the view of high unit cell volume being linked to higher ionic conductivity. It must be noted however, that the unit cell volume descriptor is not part of the optimal model's feature sets, suggesting that the underlying relationships may be more complex.

Linked to the unit cell volume and lattice parameters is the ionic radius of any introduced dopant. In Equation 4, both the ionic radii of the mobile ion and structural metal components are present and shown to be positively and negatively correlated to the output ionic conductivity, respectively. Furthermore, our cross-correlational analysis results suggest that both component ionic radii are negatively correlated to higher overall unit cell volume. Recent experimental evidence in spinel ionic compounds has suggested that the need to maximize unit cell free volume must be balanced by a minimization in strain in the host lattice.[55] Hayashi *et al.* observed that the ionic radius of any such dopant must be similar to that of the host lattice to minimize structural strain, where the similarity can be judged with a computational tolerance factor.[56]The work of Kim *et al.* and Kilner on oxygen ionic conductivity in oxide-fluorites further suggested that a critical ionic radius exists for ionic conductors that results in a lack of both expansion or contraction movements in the host lattice.[57, 58] Experimental evidence has concluded that increased dopant cation size exhibits a negative correlation with ionic conductivity, which has been explained by the steric and electronic interactions of the larger dopant cations and anionic vacancies hindering movements of the latter within the lattice.[59, 60] These results are in agreement with our resulting equation, where increasing ionic radius of the structural metal cations is negatively correlated with predicted ionic conductivity.

The number of phosphate groups has the highest negative effect on the overall ionic conductivity, which is understandable given its high Pearson correlation factor (-0.772). This correlation has been previously described in literature, and was thought to originate from the possibility of negatively charged anionic groups disturbing the pathways of conducting cations and hence reducing the rate of ionic conduction.[61] Concomitant studies by Sendek *et al.* arrived at a similar conclusion when modelling ionic electrolyte conductivities from computed electronic features, describing the ideal conduction pathway to be mostly covalent in character, where a cation can rapidly form and break bonds along the axis of travel with minimal energy penalty.[62] Recent studies by Ong et. al and Reed et. al have further linked the high electronegativity of the anionic components in solid ionic conductors with poor ionic conductivity, which was attributed to the preference for strong ionic bonding.[63, 64] Interestingly, it has been demonstrated that the strong ionic bonding observed by the anionic component leads to increased electrochemical stability, with the discrepancy best exemplified by the difference between the ionic conductivities of sulfide-based and oxide-based Li-ion conductors.[65, 66]

Finally, our model also suggests that compounds featuring structural cations with higher base electronegativity will tend to reduce the overall ionic conductivity. While a high electronegativity in the mobile cation has been shown to reduce ionic conductivity in SICON-type solids, the effect of immobile structural cation electronegativity has not been systematically explored.[67] As mentioned above, ideal conduction pathways should be covalent in character, which would naturally filter out the use of highly electronegative dopants to improve conductivity. A large difference between the electronegativities between different components will lead to a preference for ionic bonding, reducing ionic conductivity. Interestingly, a study on LISICON-type lithium titanium phosphate (LTP) by Wang *et al.* suggested that the substitution of the less electronegative Ti(III) with the more electronegative Ti(IV) would increase ionic conductivity, which was theorized to originate from the increased covalent character in the formed Li-O-Ti bond. However, as the study focused on altering the amount of lithium within the structure, and featured no electronic characterization, it remains speculative in nature. [68]

**Limitations**

Despite high training, validation, and test set accuracies, our model is not free from capability limitations. Firstly, the training dataset is small, and the low number of highly conductive examples may influence the classification accuracy of compounds of that class, although this cannot be conclusively determined without comparing the performance of our model to one trained with additional data on highly conductive NASICON compounds.

Secondly, our model's features do not capture any bonding-related or defect-related interactions or effects, relying instead on counterparts derived from isolated constituent atoms. While the structural parameters are related to the number and type of species occupying a unit cell, the features used by the best performing models with high feature count (n>6) at  do not imply any direct relationship between the formulaic number of a species in a unit cell with ionic conductivity. A positive relationship between having multiple structural cationic species and ionic conductivity has been previously documented in solid oxide electrolytes, which our current model fails to capture.[69] As a result, drastic changes in ionic conductivity as a result of increasing levels of dopants may fail to be correctly modelled.  It is important to note that the underlying mechanics and electronic interactions governing ionic conduction are complex and a regression of features obtained from chemical intuition and domain expertise may prove to be unsatisfactory to capture this complexity, especially to establish causality rather than demonstrate correlations. These can be captured by descriptors based on DFT-calculations or lattice dynamics such as the average vibrational frequency of a mobile cationic sublattice.[62, 70] However, we aimed to develop a model that would capture some information about the relative importance of different elemental and structural features derivable from structural information on the unit cell and compound constituents to the overall conductivity, while demonstrating the building of screening algorithms using constrained datasets.

In addition, while the formulaic definitions of our materials may be known, the experimental details behind each conductivity measurement are known to be inconsistent. Although ionic conductivity of solid compounds is usually measured through pelletized materials, studies have shown that the density of the generated pellets possess an effect on measured conductivities, with highly-dense sintered pellets exhibiting significantly higher conductivities than their unsintered equivalents.[60] Any systematic data collection initiative must hence maintain a consistent measurement pipeline for maximum accuracy.

Moreover, the lack of more complex features stemming from a variety of materials synthesis pathways that can affect ionic conductivity, such as the polarizability, grain size distribution, grain defect distributions, local mobile ion distributions, lattice stress, and structural restrictions, are not modelled, and it is believed that the adoption of such features would further increase the precision of our model, at the cost of more complex feature calculations.[71, 72] The correlations behind the features identified by our model may also not be causal, and may instead be capturing the importance of some other feature aspect not considered, as supported by the poor individual Pearson correlations to the ionic conductivity of individual features. Guin *et al.*'s previous work on statistically modelling NASICON ionic conductivity suggested that ionic radius of the dopants play a critical role; such a relationship was not directly observed here, but may have been captured by the prominence of the dopant electronegativity, as the two has been shown to be linked through the effective ionic potential, which is defined by both the ionic radius and the ionization energies of an element.[5, 73] As previously mentioned, a study Sendek *et al.* on lithium-based superionic conductors demonstrated a higher cross validation accuracy when utilizing more complex structural and electronic features, however the requirement of such features would decrease the value of such a tool for experimentalists[62]

## Conclusions

We have developed a simple, cost-effective screening algorithm for ionic conductivity for NASICON compounds using widely available, computationally simple elemental features. By using the limited data available in literature, we systematically evaluated feature combinations to identify the model with the highest confidence in predicting ionic conductivity in both Li- and Na- based SICON-type compounds, with a training and cross validated accuracy of over 87 and 82%, respectively, using only features derived from the formulaic unit in a facile process. Through this process, we reduced the number of features considered from 47 to 8, a reduction of approximately 83%, while simultaneously increasing the performance of our models. Our models also provide intuition behind the chemistry behind fast ion conductors. With more available training data, increased confidence and more intuitive conclusions could be achieved in the future, while the introduction of more specialized features such as anionic electronegativity could lead improved cross-domain classification beyond R3c-type conductors.

With the increase in interest in solid ionic electrolytes, the use of efficient high-performance algorithms can drastically cut down on the amount of wasted labour and reduce the time-to-market of research outputs. To better investigate the effects of structural composition and atomic arrangements on ionic conductivity, the development of classification models utilizing DFT-based features and structural features based on simulated XRD data should be pursued, for more accurate electronic and geometric representations, respectively . In the absence of the most suitable features, the introduction of a diverse and larger set of training examples would undoubtedly improve the robustness of our model. We recommend the experimental community to aggressively increase the size of the training data available in literature, with focus on the inclusion of more highly conductive examples in primary or supplementary data, as these will serve to improve classifier robustness, as well as to provide further data to evaluate the performance of our model feature selections.

## Acknowledgements

The Engineering and Physical Sciences Research Council are thanked for funding the Centre for Doctoral Training in Molecular Modelling & Materials Science (UCL, UK; EPSRC reference EP/L015862/1) and A*Star (Singapore) are thanked for supporting a studentship for YX. KH would like to thank AME Programmatic Fund by the Agency for Science, Technology and Research under Grant No. A1898b0043.

Supplementary: **Machine-learning assisted cross-domain prediction of ionic conductivity in sodium and lithium-based superionic conductors**

## S1. Motivation for the use of Logistic Regression

In our study on ionic conductivities, classification was chosen over value prediction due to the relatively small size of the training dataset and due to the risk of experimental noise, as ionic conductivity measurements on the same material can often vary by an order of magnitude or more due to sample and microstructure variability, as well as human error. By establishing multiple classes through the use of decision boundaries several orders of magnitude can be grouped together, overcoming noise or overfitting-related challenges while still providing an effective screening tool.

While logistic regression was chosen as the algorithm behind our classifier due to its performance and interpretability, several other approaches were also considered, including XGBoost classification, support vector machine classification, neural networks, and random forests. none of these methods exhibited superior performance to logistic regression during cross validation evaluation. This is primarily due to the their preference of well-distributed, large datasets as well as their inherent high-variance classification approaches, where variance is defined as  the variability of model's prediction for a given data point or a value.[1, 2] Methods utilizing neural networks and support vector machines do not provide easily interpretable models, resulting of limited use as screening classifiers, while tree-based models primarily return qualitative feature importance rankings.

Logistic regression overcomes these limitations due to it's low-variance classification approach, high adaptability to small datasets, and capability to provide easily interpreted feature coefficients, making it the most suitable approach for this study. However, should large sample sets (n>200) within the domain become available, alternative approaches should be re-evaluated.

Table S1a. List of training and test examples utilized in this study. All values were experimentally measured in the source provided. Test set examples have been bolded for clarity.

| Formula | Ionic Conductivity (S $cm^{-1}$) | Source |
|---|---|---|
| NaZr2(PO4)3 | 4.5E-06 | 17 |
| NaGe2(PO4)3 | 1.10E-12 | 18 |
| NaTi2(PO4)3 | 4.43E-10 | 18 |
| NaHf2(PO4)3 | 8.77E-10 | 18 |
| NaSn2(PO4)3 | 4.65E-09 | 17 |
| NaGe0.5Ti1.5(PO4)3 | 5.91E-13 | 17 |
| NaGeTi(PO4)3 | 8.50E-12 | 17 |
| NaGe1.5Ti0.5(PO4)3 | 3.13E-11 | 17 |
| NaSn0.5Ti1.5(PO4)3 | 1.77E-11 | 17 |
| NaSnTi(PO4)3 | 6.86E-11 | 17 |
| NaSn1.5Ti0.5(PO4)3 | 5.14E-10 | 17 |
| NaSn0.5Zr1.5(PO4)3 | 4.23E-11 | 17 |
| NaSnZr(PO4)3 | 2.47E-10 | 17 |
| NaSn1.5Zr0.5(PO4)3 | 7.91E-10 | 17 |
| NaNbZr(PO4)3 | 2.49E-08 | 19 |
| NaNbTi(PO4)3 | 1.59E-06 | 19 |
| NaMoZr(PO4)3 | 2.06E-09 | 19 |
| NaMoTi(PO4)3 | 3.07E-07 | 19 |
| Na3Zr2(SiO4)2(PO4) | 0.00067 | 16 |
| Na4Zr2(SiO4)3 | 8.87E-09 | 22 |
| Na2.4Hf2(SiO4)1.4(PO4)1.6 | 0.00073 | 32 |
| Na2.6Hf2(SiO4)1.6(PO4)1.4 | 0.00059 | 32 |
| Na2.8Hf2(SiO4)1.8(PO4)1.2 | 0.00069 | 32 |
| Na3Hf2(SiO4)2(PO4) | 0.0011 | 32 |
| Na3.2Hf2(SiO4)2.2(PO4)0.8 | 0.0023 | 32 |
| Na3.4Hf2(SiO4)2.4(PO4)0.6 | 0.0014 | 32 |
| Na3.6Hf2(SiO4)2.6(PO4)0.4 | 0.0012 | 32 |
| Na3.8Hf2(SiO4)2.8(PO4)0.2 | 0.00032 | 32 |
| Na1.2In0.2Zr1.8(PO4)3 | 2.08E-07 | 25 |
| Na1.4In0.4Zr1.6(PO4)3 | 9.76E-07 | 25 |
| Na1.5Al0.5Zr1.5(PO4)3 | 5.7E-06 | 26 |
| Na1.5Cr0.5Zr1.5(PO4)3 | 0.00001 | 26 |
| Na1.5Ga0.5Zr1.5(PO4)3 | 3.4E-06 | 26 |
| Na1.5In0.5Zr1.5(PO4)3 | 0.000029 | 26 |
| Na1.5Sc0.5Zr1.5(PO4)3 | 0.000058 | 26 |
| Na1.5Y0.5Zr1.5(PO4)3 | 0.000056 | 26 |
| Na1.5Yb0.5Zr1.5(PO4)3 | 0.00003 | 26 |
| Na2AlZr(PO4)3 | 1.2E-06 | 26 |
| Na2CrZr(PO4)3 | 0.000025 | 26 |
| Na2.2In1.2Zr0.8(PO4)3 | 2.63E-06 | 25 |
| Na2.5Cr1.5Zr0.5(PO4)3 | 0.00018 | 26 |
| Na2.5In1.5Zr0.5(PO4)3 | 0.0001 | 26 |
| Na2.5Sc1.5Zr0.5(PO4)3 | 0.00056 | 26 |
| Na2.5Y1.5Zr0.5(PO4)3 | 0.000046 | 26 |
| Na2.5Yb1.5Zr0.5(PO4)3 | 0.00019 | 26 |
| Na2.6In1.6Zr0.4(PO4)3 | 2.81E-06 | 25 |
| Na2.8In1.8Zr0.2(PO4)3 | 2.38E-06 | 25 |
| Na3Cr2(PO4)3 | 1.7E-07 | 26 |
| Na3Fe2(PO4)3 | 1.2E-07 | 26 |
| Na3Sc2(PO4)3 | 2.27E-05 | 26 |
| Na1.4Al0.4Ge1.6(PO4)3 | 7.28E-10 | 30,31 |
| Na1.4Al0.4Sn1.6(PO4)3 | 1.41E-08 | 30,31 |
| Na1.4Al0.4Ti1.6(PO4)3 | 5.6E-08 | 30,31 |
| Na1.6Al0.6Ti1.4(PO4)3 | 1.1E-07 | 30,31 |
| Na1.8Al0.8Ti1.2(PO4)3 | 1.2E-07 | 30,31 |
| Na1.9Al0.9Ti1.1(PO4)3 | 1.3E-07 | 30,31 |
| Na1.4In0.4Ti1.6(PO4)3 | 1.86E-08 | 25 |
| Na1.4In0.4Sn1.6(PO4)3 | 2.72E-08 | 25 |
| Na1.4In0.4Hf1.6(PO4)3 | 1.86E-07 | 25 |
| Na2.5Sc0.2Zr1.8(SiO4)1.3(PO4)1.7 | 0.000319 | 36 |
| Na3Sc1.5Zr0.5(SiO4)0.5(PO4)2.5 | 0.000117 | 36 |
| Na3ScZr(SiO4)2(PO4) | 0.000182 | 36 |
| Na3Sc0.8Zr1.2(SiO4)1.2(PO4)1.8 | 0.000142 | 36 |
| Na3.5Sc0.5Zr1.5(SiO4)2(PO4) | 0.000488 | 36 |
| Na2.7Sc0.2Zr1.8(SiO4)1.5(PO4)1.5 | 8.87E-05 | 36 |
| Na1.25Sn0.25Ge1.75(PO4)3 | 7.00E-06 | 33 |
| Na1.5Sn0.5Ge1.5(PO4)3 | 8.39E-05 | 33 |
| Na1.75Sn0.75Ge1.25(PO4)3 | 1.20E-05 | 33 |
| Na1.3Al0.3Zr1.7(PO4)3 | 6.30E-05 | 34 |
| Na1.3Al0.3Ti1.7(PO4)3 | 1.40E-05 | 34 |
| **Formula** | | **Source** |
| **Na3MgZr(PO4)3** | **1.00E-06** | **37** |
| **Na3MnZr(PO4)3** | **1.80E-06** | **37** |
| **Na3V2(PO4)3** | **3.00E-08** | **38** |
| **Na3V1.9Fe0.1(PO4)3** | **2.00E-06** | **38** |
| **Na3V1.9Al0.1(PO4)3** | **1.00E-07** | **38** |
| **Na3.1V1.9Ni0.1(PO4)3** | **1.20E-06** | **38** |
| **Na3V1.9Cr0.1(PO4)3** | **4.00E-08** | **38** |
| **Na1.2Zr1.8Fe0.2(PO4)3** | **2.54E-06** | **39** |
| **Na3.4Sc2(SiO4)0.4(PO4)2.6** | **4.00E-03** | **36** |
| **Na3Al2(PO4)3** | **3.00E-08** | **41** |
| **NaAlSb(PO4)3** | **2.12E-08** | **48** |
| **LiTi2(PO4)3** | **3.60E-08** | **42** |
| **Li1.3AL0.3Ti1.7(PO4)3** | **3.40E-07** | **42** |
| **LiZr2(PO4)3** | **1.00E-05** | **43** |
| **Li1.2Zr1.8Ca0.2(PO4)3** | **1.30E-05** | **43** |
| **Li1.4Zr1.6Ca0.4(PO4)3** | **8.50E-06** | **43** |
| **Li1.1Hf1.9Cr0.1(PO4)3** | **3.80E-04** | **44** |
| **LiZr1.9Sr0.1(PO4)3** | **3.44E-05** | **45** |
| **LiZr1.8Sr0.2(PO4)3** | **1.42E-05** | **45** |
| **Li2TiFe(PO4)3** | **3.34E-07** | **46** |
| **Li2ZrFe(PO4)3** | **1.14E-08** | **46** |
| **Li2ZrIn(PO4)3** | **8.22E-08** | **47** |

Table S1b. List of features used for the model evaluation process, with Pearson Correlation Factors. Features without Pearson correlation factors not considered in the model are bolded. Here, the symbols D1, D2, and D3 refer to the elements occupying the Na, M1, and M2 positions in $NaM1M2P_3O_{12}$.

| Feature | Description | Pearson Correlation Factor |
|---|---|---|
| **S** | **Ionic conductivity at room temperature (S cm$^{-2}$)** | **N/A** |
| **Slog** | **Logarithmic transformation of S** | **N/A** |
| a | Lattice parameter (Å) | 0.457 |
| c | Lattice parameter (Å) | -0.054 |
| a/c | Ratio of a to c parameters | 0.333 |
| h | Lattice a-plane hypotenuse (Å) | 0.457 |
| d | Cell diameter (Å) | 0.026 |
| Vcalc | Calculated volume according to R3c space group | 0.460 |
| **D1ionicr** | **Ionic radius of Na** | **N/A** |
| D2ionicr | Ionic radius of formulaic leftmost dopant (D2) | 0.155 |
| D3ionicr | Ionic radius of formulaic rightmost dopant (D3) | 0.243 |
| D1eff | Calculated effective ionic radius according to formula (Na) | 0.546 |
| D2eff | Calculated effective ionic radius according to formula (D2) | 0.155 |
| D3eff | Calculated effective ionic radius according to formula (D3) | -0.079 |
| **D1volperatom** | **Calculated volume of Na** | **N/A** |
| D2volperatom | Calculated volume of D2 | 0.113 |
| D3volperatom | Calculate volume of D3 | 0.242 |
| D1vol | Calculated effective volume of species D1 according to formula (Na) | 0.546 |
| D2vol | Calculated effective volume of species D2 according to formula (D2) | 0.048 |
| D3vol | Calculated effective volume of species D1 according to formula (D3) | 0.009 |
| **D1eneg** | **Electronegativity of Na** | **N/A** |
| D2eneg | Electronegativity of dopant D2 | -0.460 |
| D3eneg | Electronegativity of dopant D3 | -0.284 |
| D1eneff | Calculated effective electronegativity according to formula (Na) | 0.546 |
| D2eneff | Calculated effective electronegativity according to formula (D2) | -0.055 |
| D3eneff | Calculated effective electronegativity according to formula (D3) | -0.213 |
| D2occu | Stoichiometric occupancy of D2 dopant position | 0.134 |
| [M] | Stoichiometric number of species M according to formula, where M represent all non-anionic species in dataset. | N/A |
| [PO4] | Stoichiometric number of species PO4 according to formula | -0.772 |
| [Na] | Stoichiometric number of species Na according to formula | 0.546 |

Table S1c. Determination methodology for features described in Table S1a.

| Feature | Determination approach |
|---|---|
| S | Experimentally determined |
| Slog | = $Log_{10}(S)$ |
| a | Tabulated |
| c | Tabulated |
| a/c | $= \frac{a}{c}$ |
| h | $= \sqrt{a^2 + a^2}$ |
| d | $= \sqrt{c^2 + h^2}$ |
| Vcalc | **Calculated volume according to R3c space group** |
| D1ionicr | Tabulated |
| D2ionicr | Tabulated |
| D3ionicr | Tabulated |
| D1eff | $= D1ionicr \times N_{Na}$ |
| D2eff | $= D2ionicr \times N_{D2}$ |
| D3eff | $= D3ionicr \times N_{D3}$ |
| D1volperatom | $=\frac{4}{3}\pi(D1ionicr)^3$ |
| D2volperatom | $=\frac{4}{3}\pi(D2ionicr)^3$ |
| D3volperatom | $=\frac{4}{3}\pi(D3ionicr)^3$ |
| D1vol | $=\frac{4}{3}\pi(D1ionicr)^3 \times N_{Na}$ |
| D2vol | $=\frac{4}{3}\pi(D2ionicr)^3 \times N_{D2}$ |
| D3vol | $=\frac{4}{3}\pi(D3ionicr)^3 \times N_{D3}$ |
| D1eneg | Tabulated |
| D2eneg | Tabulated |
| D3eneg | Tabulated |
| D1eneff | $= D1eneff \times N_{Na}$ |
| D2eneff | $= D2eneff \times N_{D2}$ |
| D3eneff | $= D3eneff \times N_{D3}$ |
| D2occu | Incremented based on formulaic occupancy of D2 dopant position |
| [M] | Element-specific Incrementation based on formulaic count of species in D2 or D3 |
| [PO4] | Incremented based on formulaic count of PO4 in compound |
| [Na] | Incremented based on formulaic count of Na in compound |

Table S2. Training and CVMR accuracies of models trained with increasing feature combinations.

| Number of Features / combination | Feature combination | TMR | CVMR | Standard deviation |
|---|---|---|---|---|
| 10 | 'a', 'c', 'h', 'd', 'D1ionicr', 'D2ionicr', 'PO4', 'D1eneg', 'D2eneg', 'D3eneg' | 0.871 | 0.831 | 0.358 |
| 9 | 'a', 'c', 'h', 'd', 'D2ionicr', 'PO4', 'D1eneg', 'D2eneg', 'D3eneg' | 0.871 | 0.831 | 0.358 |
| 8 | 'a', 'c', 'h', 'd', 'D2ionicr', 'PO4', 'D2eneg', 'D3eneg' | 0.886 | 0.831 | 0.358 |
| 7 | 'c', 'h', 'd', 'D2ionicr', 'PO4', 'D2eneg', 'D3eneg' | 0.857 | 0.831 | 0.358 |
| 6 | 'a', 'a/c', 'D1ionicr', 'PO4', 'D2eneg', 'D3eneg' | 0.843 | 0.831 | 0.358 |
| 5 | 'a', 'D1ionicr', 'PO4', 'D2eneg', 'D3eneg' | 0.843 | 0.831 | 0.358 |
| 4 | 'a', 'PO4', 'D2eneg', 'D3eneg' | 0.843 | 0.831 | 0.358 |
| 3 | 'D3vol', 'Na', 'D3eneff' | 0.814 | 0.819 | 0.397 |

Table S3. Test accuracies of models trained with the feature combinations in Table S2, evaluated over 22 examples of Na and Li based SICON compounds.

| Number of Features / combination | Feature combination | Test accuracy |
|---|---|---|
| 10 | 'a', 'c', 'h', 'd', 'D1ionicr', 'D2ionicr', 'PO4', 'D1eneg', 'D2eneg', 'D3eneg' | 0.818 |
| 9 | 'a', 'c', 'h', 'd', 'D2ionicr', 'PO4', 'D1eneg', 'D2eneg', 'D3eneg' | 0.818 |
| 8 | 'a', 'c', 'h', 'd', 'D2ionicr', 'PO4', 'D2eneg', 'D3eneg' | 0.818 |
| 7 | 'c', 'h', 'd', 'D2ionicr', 'PO4', 'D2eneg', 'D3eneg' | 0.818 |
| 6 | 'a', 'a/c', 'D1ionicr', 'PO4', 'D2eneg', 'D3eneg' | 0.818 |
| 5 | 'a', 'D1ionicr', 'PO4', 'D2eneg', 'D3eneg' | 0.818 |
| 4 | 'a', 'PO4', 'D2eneg', 'D3eneg' | 0.818 |
| 3 | 'D3vol', 'Na', 'D3eneff' | 0.328 |